\documentclass[sigconf, review=false]{acmart}

\usepackage{listings}
\definecolor{Comments}{rgb}{0.00,0.50,0.00}
\definecolor{KeyWords}{rgb}{0.00,0.00,0.63}
\definecolor{Strings}{rgb}{0.84,0.00,0.00}

\lstdefinestyle{ACM}{%
  basicstyle=\scriptsize\ttfamily,%
  keywordstyle=\color{KeyWords},%
  showstringspaces=false,%
  identifierstyle=\color{black},%
  commentstyle=\color{Comments},%
  stringstyle=\color{Strings},%
  frame=shadowbox,
  rulesepcolor=\color{black},%
  numbers=left,
  firstnumber=1,%
  stepnumber=5,%
  columns=fixed,
  fontadjust=true,%
  basewidth=0.5em,%
  captionpos=t,%
  abovecaptionskip=\smallskipamount,
  belowcaptionskip=\smallskipamount,
}

\setcopyright{rightsretained}

\acmConference[PEARC18]{Practice \& Experience in Research Computing
  2018}{July 2018}{Pittsburgh, PA USA}
\acmYear{2018}
\copyrightyear{2018}

\begin{document}
\title{Jupyter as Common Technology Platform for Interactive HPC Services}
\subtitle{}

\author{Michael B. Milligan}
\affiliation{%
  \institution{Minnesota Supercomputing Institute \\ University of Minnesota}
  \streetaddress{117 Pleasant St. SE}
  \city{Minneapolis} 
  \state{MN} 
  \postcode{55455}
}
\email{milligan@umn.edu}

\renewcommand{\shortauthors}{M. Milligan}

\copyrightyear{2018}
\acmYear{2018}
\setcopyright{acmlicensed}
\acmConference[PEARC '18]{Practice and Experience in Advanced Research Computing}{July 22--26, 2018}{Pittsburgh, PA, USA}
\acmBooktitle{PEARC '18: Practice and Experience in Advanced Research Computing, July 22--26, 2018, Pittsburgh, PA, USA}
\acmPrice{15.00}
\acmDOI{10.1145/3219104.3219162}
\acmISBN{978-1-4503-6446-1/18/07}

\begin{abstract}
The Minnesota Supercomputing Institute has implemented Jupyterhub and
the Jupyter notebook server as a general-purpose point-of-entry to
interactive high performance computing services. This mode of
operation runs counter to traditional job-oriented HPC operations, but
offers significant advantages for ease-of-use, data exploration,
prototyping, and workflow development. From the user perspective,
these features bring the computing cluster nearer to parity with
emerging cloud computing options. On the other hand, retreating from
fully-scheduled, job-based resource allocation poses challenges for
resource availability and utilization efficiency, and can involve
tools and technologies outside the typical core competencies of a
supercomputing center's operations staff. MSI has attempted to
mitigate these challenges by adopting Jupyter as a common technology platform for
interactive services, capable of providing command-line, graphical,
and workflow-oriented access to HPC resources while still integrating
with job scheduling systems and using existing compute resources. This
paper will describe the mechanisms that MSI has put in place,
advantages for research and instructional uses, and lessons learned.

\end{abstract}




\maketitle

\section{Introduction}

Users of academic research computing services display a wide range of
familiarity with high-performance computing technology, but the large majority
are focused on scientific or learning outcomes rather than computational
practice\cite{Stitt:PRACE}. As a result, a significant amount of software
development effort associated with a high-performance computing center focuses
on interfaces--particularly web interfaces--enhancing the usability of computing
systems and accessibility of technical information.

Different challenges result depending on the type and location of these software
development activities. They may be undertaken by users, acting alone or in
groups, or by dedicated center staff, who may be dedicated software developers
or operational staff engaging in development activities as a secondary task. In
the case of user-driven development, center staff is challenged to either
support diverse software packages or support users attempting to deploy such
packages with the limited access permitted to unprivileged users. When
development is staff-driven, the option exists to build around common base
technologies and reduce the cognitive load on both development and support
staff. Moreover, if the staff-deployed usability technologies are sufficiently
flexible, they may displace some demand for user-selected software by
substituting domain-general platforms for domain-specific
gateways\cite{Sampedro:PEARC17}.

The Minnesota Supercomputing Institute (MSI) at the University of Minnesota has
adopted a goal of supporting Interactive HPC as a first-class service. The
availability of interactive services can provide significant benefits for data
visualization and exploration\cite{yu:vizic}, workflow
prototyping\cite{Sampedro:PEARC17}, and training\cite{mb:cornellJRS}. This mode
of operation is strongly desired by currently-existing users, who are routinely
willing to sacrifice performance (by computing on local resources) or cost (by
purchasing access to external cloud computing resources) to achieve flexibility
not offered by traditional HPC. At present MSI supports several interactive
modes of operation, including traditional command line interface (CLI) tools,
graphical remote desktop sessions, and web-based services. These features bring
MSI service offerings closer to parity with both local and emerging cloud
computing options.

In addition MSI employs a dedicated core of staff software developers, primarily
supported by grants and contracts from the research community, focused on
application development that strongly leverages the availability of
high-performance computing resources. In practice, nearly all of the development
projects supported in this way include a web application component supporting
usability or accessibility. 

In this paper, we describe efforts at MSI to provide both interactive HPC
services, and robust application development support of usability and
accessibility technologies, using components of the Jupyter software ecosystem
as a common technology platform. MSI has implemented a public JupyterHub service
that permits users to seamlessly run the interactive Jupyter Notebook web
application using normal batch-scheduled clustered computing resources. By
exploiting existing extension points in Jupyter and JupyterHub, MSI application
developers have used these components to deploy project-specific customized web
portals providing access to particular workflows and data for research and
training purposes. These efforts benefit greatly from the community-supported
open source nature of the tools in the Jupyter software ecosystem.

\section{Jupyter Components}

The Jupyter Notebook\cite{kluyver:notebooks} is a tool and a
publishing format for reproducible computational workflows. The
central concept is a notebook document, which combines formatted
textual content, executable code in one of the many supported
languages, and captured results from code execution. When opened
using the Jupyter web application, notebooks become graphical
web-oriented documents capable of recording, explaining, sharing, and
executing code, which are well-suited to capturing and reproducing
complex workflows\cite{shen:nb:nature}.  The document
format\cite{nbformat.readthedocs.io} is open and
extensible, permitting not only execution within the Jupyter Notebook
platform, but also sharing, version control, and rendering into
various web or publication formats.

The Jupyter web application currently exists in two incarnations: the
``Classic'' interface and the newer JupyterLab\cite{jupyterlab.readthedocs.io}
interface. While the latter is a more modern Single Page Application based on
PhosphorJS\footnote{\url{http://phosphorjs.github.io/}}, both provide a similar
feature set including an integrated file browser, file editing and preview,
command line terminal access, and access to all features of the Jupyter
Notebook. The web application is the extension target for web interaction
customizations, generally written in Javascript or Python, that provide new
visualization options, interactive ``widgets,'' or customizations to the
controls in the web application itself.

Jupyterhub\cite{jupyterhub.readthedocs.io} is a web application
written in Python and Node.js intended to support Jupyter notebook
servers as a multi-user service. With Jupyterhub users connect to a
configurable proxy (the Proxy) that dynamically routes requests,
either to a hub component (the Hub) that handles authentication and
spawning of notebook servers, or to a running notebook server that has
been spawned for the user. 

\subsection{Notebook server}

The architecture of the core Jupyter notebook server consists of a web
frontend application, language kernels, and a suite of standardized
protocols. The frontend application, which is accessible using any
modern web browser, implements a number of user-facing features: in
addition to display and interaction with notebook documents, it
provides file browsing, an in-browser terminal emulator for executing
shell commands, a manager for running processes, and a selector to
create new notebooks from any of the available language
kernels. Additional features may be added by supplemental modules that
extend the JavaScript code running in the browser and/or the Python
code implementing the server.

A language kernel is a module that provides the capability to execute
code in a given programming language. The kernel executes as a
separate process, and is connected to the frontend application via
network sockets. As a result the frontend is isolated from failures in
the kernel, whether due to bugs or code executed by the user. The
Jupyter project specifies the kernel communication protocol, but only
provides a reference Python kernel. The Jupyter community has created
numerous additional kernels. At the time of writing, the Jupyter
project lists 104 kernels supporting around 60 languages%
\footnote{\url{https://github.com/jupyter/jupyter/wiki/Jupyter-kernels}}.
These include kernels for many traditional interactive scripting
languages (e.g.\ Python, Perl, Bash), but also feature wrappers for
domain/application specific languages (e.g.\ MATLAB, Maxima, Gnuplot,
ROOT), interfaces to remote query execution systems such as Apache
Spark, and environments for interactive use of compiled languages
(e.g.\ C, C++, Java, assembly language).
MSI officially supports kernels for Python 2, Python 3, and R, and also enables
users to build kernels for personal use.

The Jupyter notebook server normally runs in the context of a given
user in a computing environment, providing a browser interface capable
of accessing arbitrary files and running arbitrary code. Therefore it
is important that the server runs only in the context of a user
session authorized to use computational resources, such that the
server is limited to capabilities already permitted to the user by POSIX
permissions, containerization, or other means. It is
similarly important that only the authenticated user be able to
connect to this browser interface. An unmodified installation of the
notebook server addresses these issues through the assumption of a
local user: it makes no attempt to gain or drop permissions relative
to the user invoking the application, but only permits connections
from the local machine. Access is secured via a secret token that is
displayed to the user or passed directly to the web browser.

In an institutional research computing setting, this context may
still apply in certain situations -- e.g. when a user is sitting at a
workstation in a computer lab -- but typically the effective use of
HPC resources will break this assumption in various ways. Issues
addressed by our work and described here include:

\begin{itemize}
\item The user typically is not physically proximate to the compute resource
\item The compute resource typically resides on a network with
  restricted connectivity, and in particular may not accept
  connections from any machine on which the user can easily run a web
  browser
\item The user may not have interactive shell access to the resource,
  being expected instead to submit scripts for scheduled execution
\item Institutional policy may favor or require user authentication
  through a central service, especially for web-oriented services
\item It is often possible for a user to work around the above issues,
  but the resulting procedure may be of such complexity that the ``user
  friendly'' interface is only available to highly sophisticated and
  motivated users.
\end{itemize}

\subsection{Jupyterhub}

The JupyterHub architecture addresses the issues of locality and network
topology raised above. Since the Hub and Proxy reside on a known host under the
control of the site administrator, user notebook servers may be safely
configured to accept connections from that host. Communication with the per-user
servers is secured and authenticated via an OAuth exchange between the Hub and
the notebook server. Adopting appropriate network security measures, this host
can safely bridge between the restricted internal network domain and a broader
network domain from which users can connect, potentially including the public
Internet. The Hub maintains a database of per-user state, enabling notebook
servers for different users to reside at different network addresses. The action
of the dynamic proxy allows the different hosts involved to transparently appear
to the user as a single URL.

Like the Jupyter notebook server itself, Jupyterhub is extensible,
enabling it to support a variety of usage and deployment patterns.
Two principal points of customization are the mechanisms for
authenticating users, and for spawning new user notebook
servers. These functions are accomplished by interchangeable modules
that adhere to defined APIs, and which can be selected and configured
without local code changes via Jupyterhub's configuration file. In the
default configuration, Jupyterhub uses an authenticator that simply
accepts username/password pairs for a local user, and a spawner that
simply runs notebook servers as subprocesses on the local
machine.

After a survey of the available alternative modules, MSI
decided to support development of new code enabling Jupyterhub
configurations appropriate to institutional HPC settings. Some of this
work involved improvements to existing modules (e.g.\ improvements to
Globus Auth support in the OAuth authenticator module). Other aspects,
described below, required the development of novel modules. In order to
foster software sustainability, we structured these modules
with sufficient generality to be useful to a variety
of institutions beyond MSI, with the result that they have been
formally adopted by the Jupyterhub development team, and they are used
by and receive contributions a number of users beyond MSI.

\section{Applications}

In this section we describe three key applications for which the Jupyter
technology platform has formed the basis at MSI.

\subsection{HPC Notebooks Service}

\begin{figure}
  \includegraphics[width=\linewidth]{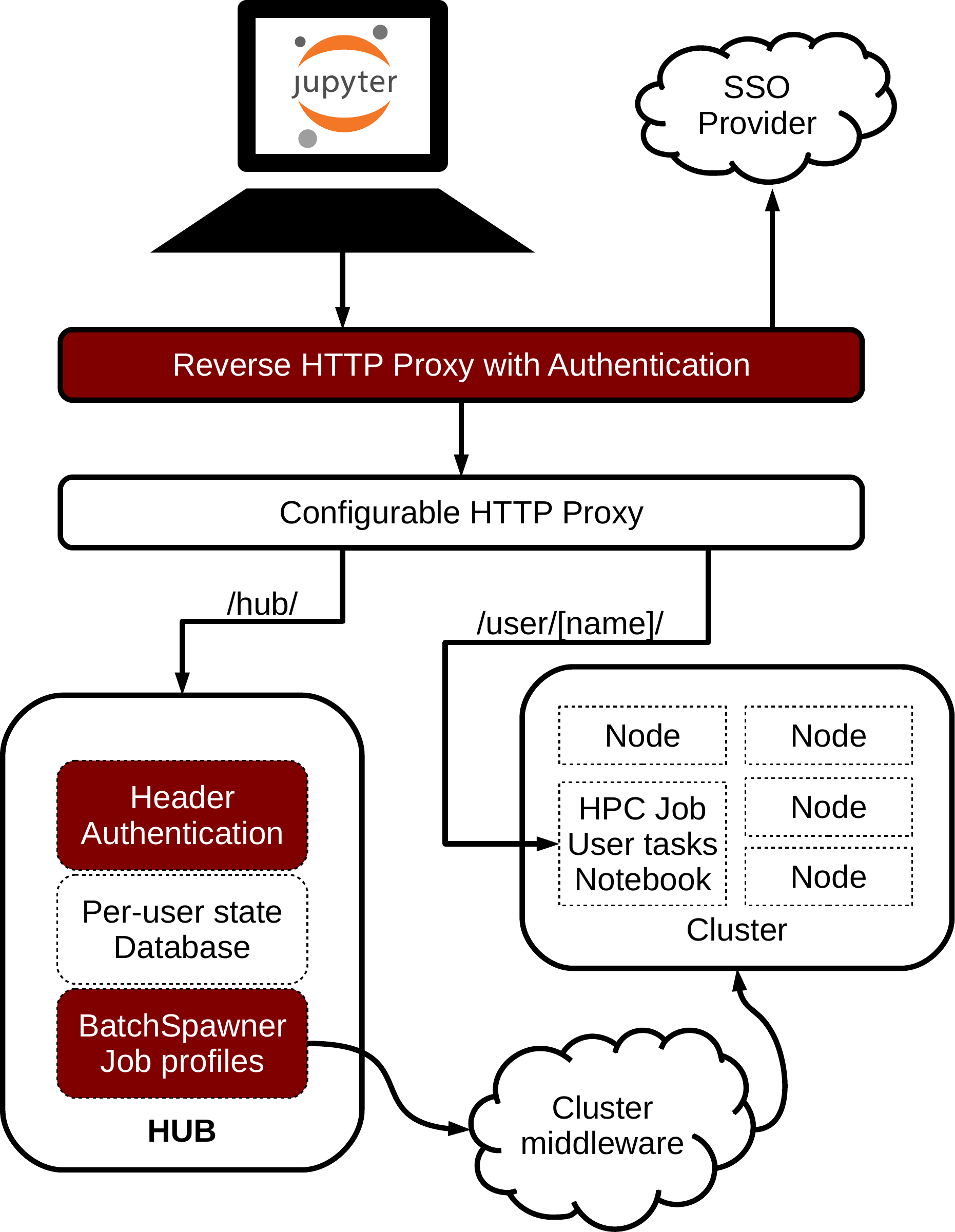}
  \caption{Architecture of the public HPC Notebooks Service implemented at MSI
    using JupyterHub. Tinted elements have been added or customized as part of
    this work.}
\end{figure}

Using the JupyterHub architecture described below, MSI has implemented a public
service that permits users to seamlessly run the interactive Jupyter Notebook
web application using normal batch-scheduled clustered computing resources.

\subsubsection{BatchSpawner}

The \texttt{LocalProcessSpawner} included with JupyterHub is a useful
default for ease of setup, but the architecture supports a range of spawner
modules for running per-user notebook servers in a variety of
contexts. The core requirements of the Spawner API are that a module
support:

\begin{itemize}
  \item \texttt{start} and \texttt{stop} methods that asynchronously
    initiate or terminate execution of a notebook server for a
    specified user,
  \item an idempotent \texttt{poll} method that yields the current
    execution state,
  \item properties for communicating e.g. the network location of the
    notebook server, and
  \item methods for rendering any process state into key-value pairs
    that become part of the persistent per-user state.
\end{itemize}

All spawner modules will thus track a network address where a user's
notebook server can be found, but each module will maintain different
types of information for managing the process. In the case of the
batch scheduling systems commonly used in HPC contexts, processes are
typically initiated as part of a scheduled job which is assigned a
persistent job identifier upon creation. The job identifier can be
used with job management commands to query job state (in particular,
whether the job has started and, if appropriate, on what host it is
running) or terminate the job. While the names and syntax of these
commands vary according to the scheduling system in use, this pattern
applies generally to such systems. Conveniently, this pattern maps
cleanly onto the requirements of the Spawner API listed above, with
the job identifier being the key item of process state that must be
retained.

We developed \texttt{BatchSpawner}%
\footnote{\url{https://github.com/jupyterhub/batchspawner}},
a Jupyterhub spawner module that uses an available batch scheduling
system to launch the user notebook server. For generality, the core
functionality is provided by a base class, \texttt{BatchSpawnerBase},
that implements the Spawner API expected by Jupyterhub, provides
facilities for templating job scripts, and provides facilities for
templating and executing job control commands. This base class does
not assume the presence of any particular scheduler, nor does it
enforce either the presence or format of parameters intended for
insertion into templates, although several commonly used parameters
(memory limit, number of cores) are provided.

Adding support in \texttt{BatchSpawner} for integration with a
specific batch scheduling system is accomplished by subclassing
the base class or one of the utility subclasses. Prior to
the initial public release, we developed support for the Torque
scheduler used at MSI and for SLURM (Simple Linux Utility for Resource
Management). By targeting two different scheduling systems during the
development process, we identified many points of overlapping
functionality and removed these to the common base classes. Adding
support for a new scheduler has proved to be simple. As an added benefit,
xadapting \texttt{BatchSpawner} to a new site can usually be accomplished
by adjusting one or more template strings via a configuration file, without
the need to create or maintain site-specific code.

Since the public release of \texttt{BatchSpawner} in 2016,
institutional users have contributed classes implementing support for
Moab, Condor, and Grid Engine-style schedulers. On the basis of this
demonstrated multi-institutional support, \texttt{BatchSpawner} has
now been adopted by the Jupyterhub developers as an
officially-supported component, thus ensuring long-term sustainability
of this software.

\begin{figure}
  \includegraphics[width=\linewidth]{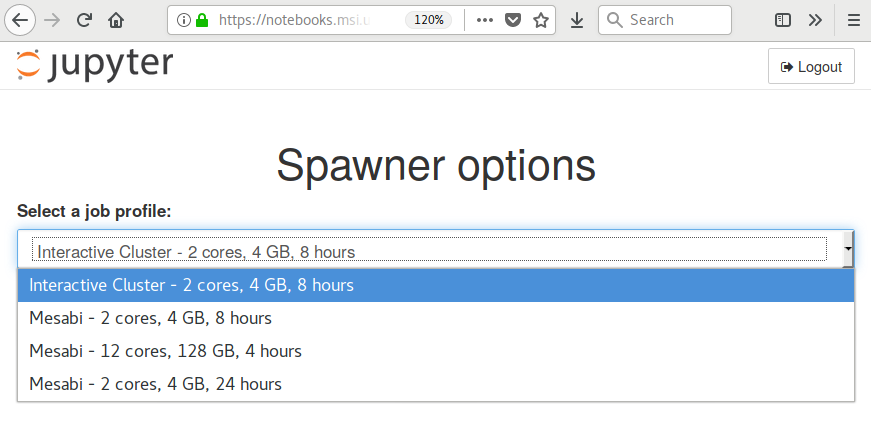}
  \caption{Interface presented to users of the HPC Notebooks Service prior to
    starting a notebook server. Options correspond to pre-selected job
    configurations specifying a target system and requested resources. The
    options correspond to \texttt{BatchSpawner} configurations, and the
    selection interface is implemented via \texttt{ProfilesSpawner}.}
\end{figure}

\subsubsection{Job Profiles}

\begin{lstlisting}[float=*th, language=Python, style=ACM,
  label={profileconf},
  caption={Representative configuration
    document for \texttt{ProfilesSpawner}, illustrating the ability to
    select spawner types and parameters. With this example
    configuration, the user can select between a local process, batch
    jobs with predefined job parameters, or a notebook server in a
    Docker container.} ]
c.ProfilesSpawner.profiles = [
    ( "Local server", 'local', 'jupyterhub.spawner.LocalProcessSpawner', {'ip':'0.0.0.0'} ),
    ('Parallel - 12 cores, 128 GB, 4 hours', 'parallel', 'batchspawner.TorqueSpawner',
          dict(req_nprocs='12', req_queue='ram256g', req_runtime='4:00:00', req_memory='125gb')),
    ('Interactive - 2 cores, 4 GB, 8 hours', 'interactive', 'batchspawner.TorqueSpawner',
          dict(req_nprocs='2', req_host='labhost.example.edu', req_queue='lab', req_runtime='8:00:00', req_memory='4gb')),
    ('Docker Python 2/3', 'docker', 'dockerspawner.SystemUserSpawner',
          dict(container_image="jupyterhub/systemuser"))
]
\end{lstlisting}

Jupyterhub implements application configuration at runtime using
\texttt{traitlets}\footnote{\url{https://traitlets.readthedocs.io/}},
a Python library that provides a convenient mechanism to populate
class properties with values found in a configuration file or other
source of configuration data. Since Python is a dynamic language, this
extends to run-time selection of the classes used to implement
features, such as selection of the \texttt{BatchSpawner}
implementation of the Spawner API. Using this mechanism, site
administrators can select BatchSpawner, and simultaneously configure
it with parameters appropriate to their site. However, a drawback of
this facility is that parameters are injected once, when the
configuration data is parsed during Hub startup. We found it desirable
to provide MSI users with the ability to select some parameters
(e.g. memory, runtime) at the time they request the spawner to start
their notebook server.

The Spawner API provides a mechanism intended for this purpose. If the
spawner provides an HTML form snippet, the Hub will display a Spawner
Options Form containing the snippet to the user before starting their
notebook server. The Spawner API passes the returned data to the
spawner, but is silent regarding how the data should be used. We
decided to extend upon this mechanism with the following goals: (1)
provide an option mechanism permitting selection of spawner type and
properties from an administrator-controlled list (2) without requiring
the administrator to maintain site-specific code and (3) without
exposing templated commands to unsanitized user input.

Goal (1) included support for additional spawning mechanisms, such as
a Docker swarm running in a cloud resource. However, this goal created
a corollary: since there is no standard interpretation of the Options
Form data, and few existing Spawners support the Options Form in any
capacity anyway, the option mechanism needed to be transparent to the
spawner class to ensure compatibility with arbitrary Spawners.

The resulting module, \texttt{WrapSpawner}%
\footnote{\url{https://github.com/jupyterhub/wrapspawner}}, simply
wraps another spawner class and proxies API calls through to the
per-user instance once a notebook server is created. The instance
configuration mechanism provided by \texttt{traitlets} is used to
inject configuration data computed at runtime. As a result, the
identity and configuration data of the wrapped class becomes part of
the per-user state, rather than globally-configured parameters of the
server.

We further developed \texttt{ProfilesSpawner}\footnote{distributed
  together with \texttt{WrapSpawner}} to create a user interface for
this capability. This spawner module accepts as its configuration a
list of profiles (sample given in Listing~\ref{profileconf}), where a
profile consists of a spawner class and a set of configuration
parameters for that class. When a user requests the creation of a
notebook server, \texttt{ProfilesSpawner} dynamically creates a
a dropdown selection box, by which the user configures their spawner with
the set of options corresponding to that profile. At MSI, we use
\texttt{ProfilesSpawner} to offer users a choice of four job types
corresponding to a default type or types that represent a
prioritization of performance, number of CPUs, runtime, or available
memory.
By controlling the allowable combinations of options,
we are able to size all of these job types experimentally to ensure
that the typical user experience falls within our guidelines for
interactive HPC.
  
Like \texttt{BatchSpawner}, \texttt{ProfilesSpawner} has been adopted
as an officially supported component by the Jupyterhub developers,
thus ensuring software sustainability.

\subsubsection{Authentication}

Jupyterhub supports swappable authentication modules via an
Authenticator API. In addition to the local user authenticator, the
Jupyterhub developers support authentication modules that integrate
with an LDAP database or OAuth credentials. Using either of these
options would have required additional development that would be
specific to MSI. On the other hand, our system administration staff
already had the ability to deploy an Apache web server configured to
integrate with the web single sign-on (SSO) mechanism common to other
MSI services. This has the advantage of transparently providing access
to the security and features of a central sign-on service, including
optional use of advanced features such as two-factor authentication
(2FA). Running Jupyterhub behind an Apache (or other) web server in
reverse proxy mode confers additional security advantages, since it
allows us to make the externally-facing network service a
well-understood server for which our system administrators are willing
and able to provide security configuration.

The third-party \texttt{REMOTE\_USER} Authenticator%
\footnote{\url{https://github.com/cwaldbieser/jhub_remote_user_authenticator}}
module partially meets this need. This module watches for a specially
named header in the request proxied through the web server, and uses
the data portion of that header as a username string. The web server
is configured to redirect unauthenticated requests to the SSO
mechanism, and for authenticated requests will populate the
appropriate request header. Since the user is considered authenticated
as soon as this header is presented, the Hub component never displays
a login prompt and the user interacts only with our institutional
authentication platform. In the future we plan to extend this mechanism
to permit use of additional user metadata supplied by the platform.

\subsection{Science Portal}

MSI is a primary location of software development and deployment for the GEMS
platform for agroinformatics data sharing and
analysis\cite{Gustafson:PEARC17}. The JupyterHub architecture forms the common
technology platform underlying the web interface and data analysis components of
the GEMS application. Use of this approach was readily accepted by the GEMS
leadership in part because Jupyter has already been used many times in similar
contexts\cite{Yin:PEARC17,perrin:webbpsf,fahk:jupyterhubESS}.

GEMS is intended to be used by a broad audience, but requires reliable identity
services for users to meet the authentication requirements of some data
providers. This is achieved using a custom JupyterHub Authenticator module,
which is closely derived from the Globus\footnote{\url{https://globus.org/}}
module provided with the \texttt{OAuthenticator} extension. Thus, with minor
software development effort, GEMS gained access to the Globus Auth federated
identity and access management (IAM) service, which is able to broker
authentication interactions with many educational and research institutions,
major cloud service providers, and fallback account creation options.

As the GEMS platform developed, data analysis tasks have been consistently
developed and executed within Jupyter notebooks, even as the underlying
computational infrastructure evolved to include local execution, Docker
containers, and dispatched execution in Spark and Kubernetes cluster
resources. The modular nature of the JupyterHub Spawner architecture allowed
GEMS developers to target new infrastructure with low effort, while the common
Jupyter Notebook format ensured that developed analysis workflows were preserved
and able to execute as-is or with adaptation in new environments.

A later goal on the GEMS feature roadmap included providing users with
the option to access additional analytics environments, since not all
users will prefer to use Jupyter notebooks. We were able to take
advantage of the distributed Jupyterhub architecture to provide this
capability earlier than planned, while retaining the authentication
and spawning mechanisms that had already been built around
Jupyterhub. We used the
nbserverproxy\footnote{\url{https://github.com/jupyterhub/nbserverproxy/}}
notebook server plugin, which enables the notebook server to proxy
arbitrary HTTP connections to other services on the same host, such
that they can be accessed at a user-visible URL with access controlled
as usual by Jupyterhub. When we discovered this feature, a notebook
extension had already been developed to permit users to launch and
connect to an RStudio
session\footnote{\url{https://github.com/jupyterhub/nbrsessionproxy}}. We
assisted in the development of an analogous plugin allowing users to
launch a graphical desktop session and use GUI applications inside a VNC
server\footnote{\url{https://github.com/ryanlovett/nbnovnc/}}, which
is then rendered in the browser using a Javascript VNC client. Both of
these options were then integrated into the GEMS demonstration environment.

\subsection{Binders for Training and Reproducibility}

Binder\cite{zero-to-jupyterhub} is an outgrowth of efforts to deploy Jupyter
automatically in containerized environments using Kubernetes and provide a
publicly accessible demonstration and training environment for Jupyter
applications\cite{Holdgraf:PEARC17}. This system combines JupyterHub with an
application (\texttt{repo2docker}) that can automatically build Jupyter-ready
Docker containers from online repositories tagged with appropriate metadata, and
a component (\texttt{BinderHub}) for automatically launching such containers
within a Kubernetes cluster. A public demonstration instance of this service is
already available at \url{https://mybinder.org/}, but as a public service
running on donated cloud resources, the available computational capacity is
necessarily limited.

As described below, MSI has implemented temporary, private BinderHub
services using public cloud resources.

\section{Experience}

MSI has implemented a Jupyter Notebooks Service, accessible at
\texttt{notebooks.msi.umn.edu} to authenticated MSI users. The first public
preview release of the service was in April 2016. Between April 2016 and
February 2017 the service was used to start 441 sessions for 113 users. In that
time, 18 users launched a notebook server at least five times. The average wait
time between request and notebook server startup was 39 seconds. Since that
time, the average utilization has risen to exceed ten sessions per day, with
over a dozen active users in a typical week.

Each semester MSI offers a workshop on using Python in scientific
computing. Starting in May 2016 and continuing through the present, the Jupyter
Notebooks Service was used as the standard platform for the workshop. Our
experience has been that time spent troubleshooting user connection issues has
been significantly reduced, and the number of users successfully reproducing the
examples consequently increased. Based on feedback surveys, students strongly
prefer the Jupyter platform over the previous console-based instruction. Many of
the 95 users who have launched a notebook four times or fewer were workshop
attendees.

In January 2018, MSI contributed a temporary private BinderHub service to a
training event (Gopher Day of Data II). The service allowed both presenters and
attendees to access demonstrations in identical environments, without any
requirement for local software installation or configuration, with full
isolation between user sessions. At peak, approximately sixty attendees were
able to simultaneously load data analysis demonstrations, with resource
utilization remaining well within the bounds of the four-node Kubernetes cluster
provisioned for the occasion. 

\begin{acks}
  The author would like to thank the developers of Jupyter and Jupyterhub,
  both for their efforts creating an exceedingly useful software platform,
  as well as for their technical assistance in implementing the work described
  here.

  Thanks are also due to the users of \texttt{BatchSpawner} and
  \texttt{WrapSpawner} who contributed issue reports, bux fixes, and new
  functionality; to the organizers and attendees of Gopher Day of Data II, and
  to the attendees of MSI training events, for providing testing and feedback
  about the systems discussed.

\end{acks}

\bibliographystyle{ACM-Reference-Format}
\bibliography{refs} 


\begin{thebibliography}{00}


\ifx \showCODEN    \undefined \def \showCODEN     #1{\unskip}     \fi
\ifx \showDOI      \undefined \def \showDOI       #1{{\tt DOI:}\penalty0{#1}\ }
  \fi
\ifx \showISBNx    \undefined \def \showISBNx     #1{\unskip}     \fi
\ifx \showISBNxiii \undefined \def \showISBNxiii  #1{\unskip}     \fi
\ifx \showISSN     \undefined \def \showISSN      #1{\unskip}     \fi
\ifx \showLCCN     \undefined \def \showLCCN      #1{\unskip}     \fi
\ifx \shownote     \undefined \def \shownote      #1{#1}          \fi
\ifx \showarticletitle \undefined \def \showarticletitle #1{#1}   \fi
\ifx \showURL      \undefined \def \showURL       #1{#1}          \fi
\providecommand\bibfield[2]{#2}
\providecommand\bibinfo[2]{#2}
\providecommand\natexlab[1]{#1}
\providecommand\showeprint[2][]{arXiv:#2}

\bibitem[\protect\citeauthoryear{Fern{\'a}ndez, Andersson, Hagenrud, Korhonen,
  Laface, and Zupanc}{Fern{\'a}ndez et~al\mbox{.}}{2016}]%
        {fahk:jupyterhubESS}
\bibfield{author}{\bibinfo{person}{Leandro Fern{\'a}ndez},
  \bibinfo{person}{Riccard Andersson}, \bibinfo{person}{Hakan Hagenrud},
  \bibinfo{person}{Timo Korhonen}, \bibinfo{person}{Emanuele Laface}, {and}
  \bibinfo{person}{Bla Zupanc}.} \bibinfo{year}{2016}\natexlab{}.
\newblock \showarticletitle{Jupyterhub at the ESS. An Interactive Python
  Computing Environment for Scientists and Engineers}. In
  \bibinfo{booktitle}{{\em 7th International Particle Accelerator Conference
  (IPAC'16), Busan, Korea, May 8-13, 2016}}. \bibinfo{publisher}{JACOW},
  \bibinfo{address}{Geneva, Switzerland}, \bibinfo{pages}{2778--2780}.
\newblock


\bibitem[\protect\citeauthoryear{Gustafson, Erdmann, Milligan, Onsongo, Pardey,
  Prather, Silverstein, Wilgenbusch, and Zhang}{Gustafson
  et~al\mbox{.}}{2017}]%
        {Gustafson:PEARC17}
\bibfield{author}{\bibinfo{person}{Andrew Gustafson}, \bibinfo{person}{Jesse
  Erdmann}, \bibinfo{person}{Michael Milligan}, \bibinfo{person}{Getiria
  Onsongo}, \bibinfo{person}{Philip Pardey}, \bibinfo{person}{Tom Prather},
  \bibinfo{person}{Kevin Silverstein}, \bibinfo{person}{James Wilgenbusch},
  {and} \bibinfo{person}{Ying Zhang}.} \bibinfo{year}{2017}\natexlab{}.
\newblock \showarticletitle{A Platform for Computationally Advanced
  Collaborative AgroInformatics Data Discovery and Analysis}. In
  \bibinfo{booktitle}{{\em Proceedings of the Practice and Experience in
  Advanced Research Computing 2017 on Sustainability, Success and Impact}} {\em
  (\bibinfo{series}{PEARC17})}. \bibinfo{publisher}{ACM}, \bibinfo{address}{New
  York, NY, USA}, Article \bibinfo{articleno}{2}, \bibinfo{numpages}{4}~pages.
\newblock
\showISBNx{978-1-4503-5272-7}
\showDOI{%
\url{https://doi.org/10.1145/3093338.3093376}}


\bibitem[\protect\citeauthoryear{Holdgraf, Culich, Rokem, Deniz, Alegro, and
  Ushizima}{Holdgraf et~al\mbox{.}}{2017}]%
        {Holdgraf:PEARC17}
\bibfield{author}{\bibinfo{person}{Chris Holdgraf}, \bibinfo{person}{Aaron
  Culich}, \bibinfo{person}{Ariel Rokem}, \bibinfo{person}{Fatma Deniz},
  \bibinfo{person}{Maryana Alegro}, {and} \bibinfo{person}{Dani Ushizima}.}
  \bibinfo{year}{2017}\natexlab{}.
\newblock \showarticletitle{Portable Learning Environments for Hands-On
  Computational Instruction: Using Container- and Cloud-Based Technology to
  Teach Data Science}. In \bibinfo{booktitle}{{\em Proceedings of the Practice
  and Experience in Advanced Research Computing 2017 on Sustainability, Success
  and Impact}} {\em (\bibinfo{series}{PEARC17})}. \bibinfo{publisher}{ACM},
  \bibinfo{address}{New York, NY, USA}, Article \bibinfo{articleno}{32},
  \bibinfo{numpages}{9}~pages.
\newblock
\showISBNx{978-1-4503-5272-7}
\showDOI{%
\url{https://doi.org/10.1145/3093338.3093370}}


\bibitem[\protect\citeauthoryear{Kluyver, Ragan-Kelley, P{\'e}rez, Granger,
  Bussonnier, Frederic, Kelley, Hamrick, Grout, Corlay, et~al\mbox{.}}{Kluyver
  et~al\mbox{.}}{2016}]%
        {kluyver:notebooks}
\bibfield{author}{\bibinfo{person}{Thomas Kluyver}, \bibinfo{person}{Benjamin
  Ragan-Kelley}, \bibinfo{person}{Fernando P{\'e}rez}, \bibinfo{person}{Brian
  Granger}, \bibinfo{person}{Matthias Bussonnier}, \bibinfo{person}{Jonathan
  Frederic}, \bibinfo{person}{Kyle Kelley}, \bibinfo{person}{Jessica Hamrick},
  \bibinfo{person}{Jason Grout}, \bibinfo{person}{Sylvain Corlay}, {and}
  \bibinfo{person}{others}.} \bibinfo{year}{2016}\natexlab{}.
\newblock \showarticletitle{Jupyter Notebooks--a publishing format for
  reproducible computational workflows}. In \bibinfo{booktitle}{{\em
  Positioning and Power in Academic Publishing: Players, Agents and Agendas:
  Proceedings of the 20th International Conference on Electronic Publishing}}.
  \bibinfo{publisher}{IOS Press}, \bibinfo{address}{Amsterdam, The
  Netherlands}, \bibinfo{pages}{87--90}.
\newblock


\bibitem[\protect\citeauthoryear{Mehringer and Birkland}{Mehringer and
  Birkland}{2015}]%
        {mb:cornellJRS}
\bibfield{author}{\bibinfo{person}{Susan Mehringer} {and}
  \bibinfo{person}{Aaron Birkland}.} \bibinfo{year}{2015}\natexlab{}.
\newblock \showarticletitle{Incorporating interactive compute environments into
  web-based training materials using the Cornell job runner service}. In
  \bibinfo{booktitle}{{\em Proceedings of the 2015 XSEDE Conference: Scientific
  Advancements Enabled by Enhanced Cyberinfrastructure}}. ACM,
  \bibinfo{pages}{20}.
\newblock


\bibitem[\protect\citeauthoryear{Perrin, Sivaramakrishnan, Lajoie, Elliott,
  Pueyo, Ravindranath, and Albert}{Perrin et~al\mbox{.}}{2014}]%
        {perrin:webbpsf}
\bibfield{author}{\bibinfo{person}{Marshall~D Perrin}, \bibinfo{person}{Anand
  Sivaramakrishnan}, \bibinfo{person}{Charles-Philippe Lajoie},
  \bibinfo{person}{Erin Elliott}, \bibinfo{person}{Laurent Pueyo},
  \bibinfo{person}{Swara Ravindranath}, {and} \bibinfo{person}{Lo{\"\i}c
  Albert}.} \bibinfo{year}{2014}\natexlab{}.
\newblock \showarticletitle{Updated point spread function simulations for JWST
  with WebbPSF}. In \bibinfo{booktitle}{{\em SPIE Astronomical Telescopes and
  Instrumentation}}, Vol.~\bibinfo{volume}{9143}.
  \bibinfo{publisher}{International Society for Optics and Photonics},
  \bibinfo{pages}{91433X--91433X}.
\newblock
\showDOI{%
\url{https://doi.org/10.1117/12.2056689}}


\bibitem[\protect\citeauthoryear{Sampedro, Hauser, and Sood}{Sampedro
  et~al\mbox{.}}{2017}]%
        {Sampedro:PEARC17}
\bibfield{author}{\bibinfo{person}{Zebula Sampedro}, \bibinfo{person}{Thomas
  Hauser}, {and} \bibinfo{person}{Saurabh Sood}.}
  \bibinfo{year}{2017}\natexlab{}.
\newblock \showarticletitle{Sandstone HPC: A Domain-General Gateway for New HPC
  Users}. In \bibinfo{booktitle}{{\em Proceedings of the Practice and
  Experience in Advanced Research Computing 2017 on Sustainability, Success and
  Impact}} {\em (\bibinfo{series}{PEARC17})}. \bibinfo{publisher}{ACM},
  \bibinfo{address}{New York, NY, USA}, Article \bibinfo{articleno}{33},
  \bibinfo{numpages}{7}~pages.
\newblock
\showISBNx{978-1-4503-5272-7}
\showDOI{%
\url{https://doi.org/10.1145/3093338.3093360}}


\bibitem[\protect\citeauthoryear{Shen}{Shen}{2014}]%
        {shen:nb:nature}
\bibfield{author}{\bibinfo{person}{Helen Shen}.}
  \bibinfo{year}{2014}\natexlab{}.
\newblock \showarticletitle{Interactive notebooks: Sharing the code}.
\newblock \bibinfo{journal}{{\em Nature\/}} \bibinfo{volume}{515},
  \bibinfo{number}{7525} (\bibinfo{year}{2014}), \bibinfo{pages}{151}.
\newblock


\bibitem[\protect\citeauthoryear{Stitt and Robinson}{Stitt and
  Robinson}{2008}]%
        {Stitt:PRACE}
\bibfield{author}{\bibinfo{person}{Tim Stitt} {and} \bibinfo{person}{Tim
  Robinson}.} \bibinfo{year}{2008}\natexlab{}.
\newblock \bibinfo{booktitle}{{\em A Survey on Training and Education Needs for
  Petascale Computing}}.
\newblock \bibinfo{type}{{T}echnical {R}eport}.
  \bibinfo{institution}{Partnership for Advanced Computing in Europe}.
\newblock
\showURL{%
\url{http://www.prace-project.eu/IMG/pdf/D3-3-1_document_final.pdf}}


\bibitem[\protect\citeauthoryear{Team}{Team}{2015}]%
        {nbformat.readthedocs.io}
\bibfield{author}{\bibinfo{person}{Project~Jupyter Team}.}
  \bibinfo{year}{2015}\natexlab{}.
\newblock \bibinfo{title}{The Jupyter Notebook Format}.
\newblock   (\bibinfo{year}{2015}).
\newblock
\showURL{%
Retrieved June 2, 2017 from \url{https://nbformat.readthedocs.io/}}


\bibitem[\protect\citeauthoryear{Team}{Team}{2018a}]%
        {jupyterhub.readthedocs.io}
\bibfield{author}{\bibinfo{person}{Project~Jupyter Team}.}
  \bibinfo{year}{2018}\natexlab{a}.
\newblock \bibinfo{title}{JupyterHub}.
\newblock   (\bibinfo{year}{2018}).
\newblock
\showURL{%
Retrieved March 24, 2018 from \url{https://jupyterhub.readthedocs.io/}}


\bibitem[\protect\citeauthoryear{Team}{Team}{2018b}]%
        {jupyterlab.readthedocs.io}
\bibfield{author}{\bibinfo{person}{Project~Jupyter Team}.}
  \bibinfo{year}{2018}\natexlab{b}.
\newblock \bibinfo{title}{JupyterLab}.
\newblock   (\bibinfo{year}{2018}).
\newblock
\showURL{%
Retrieved March 24, 2018 from \url{https://jupyterlab.readthedocs.io/}}


\bibitem[\protect\citeauthoryear{Team}{Team}{2018c}]%
        {zero-to-jupyterhub}
\bibfield{author}{\bibinfo{person}{Project~Jupyter Team}.}
  \bibinfo{year}{2018}\natexlab{c}.
\newblock \bibinfo{title}{Zero to JupyterHub with Kubernetes}.
\newblock   (\bibinfo{year}{2018}).
\newblock
\showURL{%
Retrieved March 25, 2018 from \url{https://zero-to-jupyterhub.readthedocs.io/}}


\bibitem[\protect\citeauthoryear{Yin, Liu, Padmanabhan, Terstriep, Rush, and
  Wang}{Yin et~al\mbox{.}}{2017}]%
        {Yin:PEARC17}
\bibfield{author}{\bibinfo{person}{Dandong Yin}, \bibinfo{person}{Yan Liu},
  \bibinfo{person}{Anand Padmanabhan}, \bibinfo{person}{Jeff Terstriep},
  \bibinfo{person}{Johnathan Rush}, {and} \bibinfo{person}{Shaowen Wang}.}
  \bibinfo{year}{2017}\natexlab{}.
\newblock \showarticletitle{A CyberGIS-Jupyter Framework for Geospatial
  Analytics at Scale}. In \bibinfo{booktitle}{{\em Proceedings of the Practice
  and Experience in Advanced Research Computing 2017 on Sustainability, Success
  and Impact}} {\em (\bibinfo{series}{PEARC17})}. \bibinfo{publisher}{ACM},
  \bibinfo{address}{New York, NY, USA}, Article \bibinfo{articleno}{18},
  \bibinfo{numpages}{8}~pages.
\newblock
\showISBNx{978-1-4503-5272-7}
\showDOI{%
\url{https://doi.org/10.1145/3093338.3093378}}


\bibitem[\protect\citeauthoryear{Yu, Kind, and Brunner}{Yu
  et~al\mbox{.}}{2017}]%
        {yu:vizic}
\bibfield{author}{\bibinfo{person}{Weixiang Yu}, \bibinfo{person}{M~Carrasco
  Kind}, {and} \bibinfo{person}{Robert~J Brunner}.}
  \bibinfo{year}{2017}\natexlab{}.
\newblock \showarticletitle{Vizic: A Jupyter-based interactive visualization
  tool for astronomical catalogs}.
\newblock \bibinfo{journal}{{\em Astronomy and Computing\/}}
  \bibinfo{volume}{20} (\bibinfo{year}{2017}), \bibinfo{pages}{128--139}.
\newblock
\showeprint{arXiv preprint arXiv:1701.01222}


\end{thebibliography}

\end{document}